\documentclass[prl,aps,twocolumn,groupedaddress,floats,showpacs,final,superscriptaddress]{revtex4-2}

\usepackage[utf8]{inputenc}
\usepackage[makeroom]{cancel}
\usepackage{graphicx}
\usepackage{amsmath}
\usepackage{amsfonts}
\usepackage{amssymb}
\usepackage{chngcntr}
\usepackage{color}
\usepackage[left=2cm,right=2cm,top=2cm,bottom=2cm]{geometry}

\usepackage{dcolumn}
\usepackage{bm}
\usepackage{simplewick}
\usepackage{array}
\usepackage{appendix}
\usepackage[normalem]{ulem} 
\RequirePackage[
   hyperindex,colorlinks,bookmarksnumbered,
   plainpages=true,pdfstartview=FitH]{hyperref}
\hypersetup{linkcolor=blue,urlcolor=blue,citecolor=blue}

\definecolor{purple}{rgb}{0.5,0,0.6}

\newcommand{\bp}{\textbf{p}}

\begin{document}

\title{Anomalous acoustoelectric signatures of chiral superconductivity}




\author{A.~N.~Osipov}
\affiliation{Technion -- Israel Institute of Technology, Haifa, 3200003, Israel}
\affiliation{Guangdong Technion -- Israel Institute of Technology, 241 Daxue Road, Shantou, Guangdong, China, 515063}

\author{V.~N.~Ivanova}
\affiliation{Guangdong Technion -- Israel Institute of Technology, 241 Daxue Road, Shantou, Guangdong, China, 515063}


\author{V.~M.~Kovalev}
\affiliation{Rzhanov Institute of Semiconductor Physics, Siberian Branch\\ of Russian Academy of Science, Novosibirsk 630090, Russia}
\affiliation{Novosibirsk State Technical University, Novosibirsk 630073, Russia}

\author{I.~G.~Savenko}
\email[ivan.g.savenko@gmail.com]{}
\affiliation{Guangdong Technion -- Israel Institute of Technology, 241 Daxue Road, Shantou, Guangdong, China, 515063}
\affiliation{Technion -- Israel Institute of Technology, Haifa, 3200003, Israel}

\date{\today}

\begin{abstract}
The identification of unconventional pairing in two-dimensional materials is a central challenge in modern condensed matter physics. 
While chiral $p$-wave superconductivity offers a promising platform for topological quantum computing, its detection remains elusive due to the inherent limitations of optical probes in the two-dimensional limit. 
We propose the anomalous acoustoelectric effect as a robust, alternative to optical signature of $p$-wave paring symmetry. 
We demonstrate that an acoustic wave induces a transverse dc current resulting in a measurable condensate phase difference on sample boundaries originating from the anisotropic scattering of quasiparticles in the absence of an external magnetic field. 
Crucially, the quasiparticle-mediated acoustoelectric response dominates near the critical temperature and, unlike the superconducting condensate, is not suppressed by electron-hole asymmetry factor. 
These results establish the anomalous acoustoelectric effect as a high-sensitivity electrical probe of the chiral order parameter, providing a tool for experimental detecting of unconventional pairing in superconductors.
\end{abstract}


\maketitle


Although superconductivity was discovered more than a century ago~\cite{onnes1911superconductivity}, it continues to attract significant attention from the modern scientific community. Today, superconductors are critical components in applications ranging from MRI scanners to particle accelerators, and are considered one of the most promising platforms for quantum computing~\cite{seidel2015applied}. 
Recent discoveries of high-temperature superconductivity in new materials~\cite{schilling1993superconductivity, eremets2008superconductivity, zhou2021high}, unconventional superconductivity in twisted bilayer graphene~\cite{torma2022superconductivity, cao2018unconventional}, and peculiar superconducting (SC) phenomena in MoS$_2$~\cite{lu2015evidence, costanzo2016gate, fang2018structure} have fueled intense interest in the mechanisms underlying these phases.

A central quest in this field is the identification of unconventional pairing, particularly $p$-wave superconductivity, which stands out as a primary candidate for realizing topological quantum computing via Majorana bound states~\cite{PhysRevX.6.031016}. 
Historically, the pursuit of triplet pairing was ignited by the discovery of superfluid He and further catalyzed by the exploration of heavy-fermion systems and the longstanding debate over the order parameter in Sr$_2$RuO$_4$~\cite{maeno2024thirty}. Typically, collecting indirect evidence for the properties of a SC pairing requires complex and delicate experiments, including phase-sensitive measurements~\cite{nelson2004odd}, nuclear magnetic resonance measurements of the Knight shift~\cite{ishida1998spin}, $\mu$SR measurements~\cite{luke1998time}, and measurements of the polar Kerr effect~\cite{xia2006high}. The difficulty of such measurements means that for many modern unconventional superconductors, the pairing symmetry remains an open question. This is exemplified by Sr$_2$RuO$_4$, where a broad consensus on chiral $p$-wave pairing~\cite{nelson2004odd, ishida1998spin, luke1998time, xia2006high, maeno2011evaluation} was dramatically overturned by recent, more precise experiments that provided unambiguous evidence for an even-parity state~\cite{pustogow2019constraints, chronister2021evidence, petsch2020reduction, maeno2024thirty}. The frontier has now shifted toward 2D (2D) transition metal dichalcogenides, such as MoS$_2$, where the interplay of broken inversion symmetry and strong spin-orbit coupling creates a unique laboratory for exotic intravalley $p$-wave pairing.

In this Letter, we propose the anomalous acoustoelectric effect as a robust, non-optical probe for detecting unconventional pairing symmetry, specifically intravalley $p$-wave states. While it is known that chiral $p$-wave superconductors can host anomalous transport phenomena~\cite{PhysRevB.80.104508, PhysRevB.92.100506, tanaka2023nonlinear}, traditional optical probes in the 2D limit often suffer from low signal-to-noise ratios and significant substrate interference. 
Furthermore, recent theoretical analyses of the photon drag effect based on Ginzburg-Landau equations have demonstrated a condensate response, proportional to electron-hole asymmetry~\cite{PhysRevLett.132.096001, cr1g-wnmc}, which usually weakens the response.

Probing material properties via surface acoustic waves (SAWs) offers important advantages. 
These include relatively simple generation in piezoelectric heterostructures and the ability to impose a low-frequency perturbation that drags excitations~\cite{parmenter1953acousto}. 
The nonlinear acoustoelectric effect, which appears at second order in perturbation theory, produces a time-independent current that can be measured using simple electronic devices. The acoustoelectric current reflects key properties of the system, including its symmetry and topology~\cite{PhysRevLett.122.256801, li2014absorption, esslinger1992acoustoelectric}, deformation induced pseudogauge fields~\cite{PhysRevLett.124.126602, PhysRevB.105.125407} and transport characteristics~\cite{miseikis2012acoustically, kovalev2015effect}. 
Thus, the acoustic waves provide a valuable tool for characterizing the physical properties of electrons in low-dimensional systems with increasing impact in modern physics~\cite{cao2025phase, su2025quantum, gunnink2026surface, song2026acoustoelectric, sano2024acoustomagnonic}. 

\begin{figure}[!t] \includegraphics[width=0.99\columnwidth]{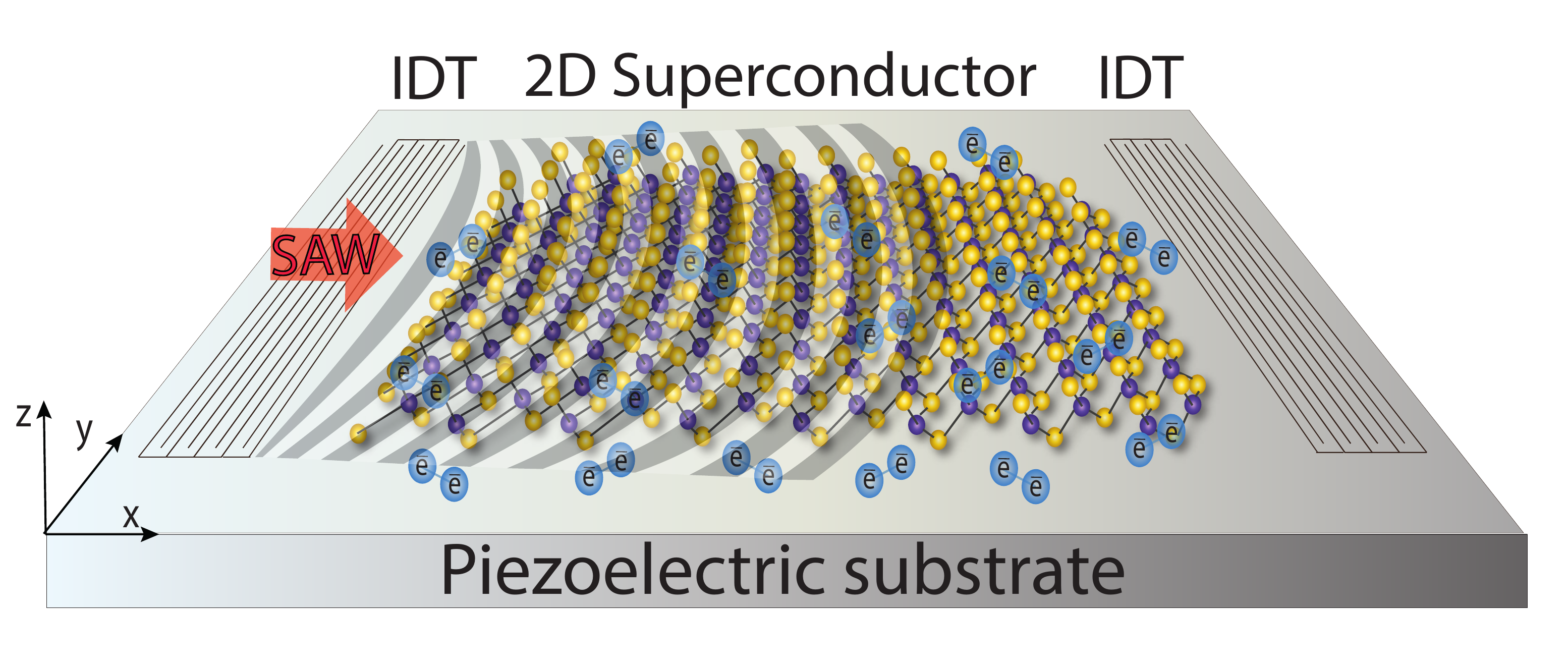} 
\caption{Schematic illustration of the system: A 2D superconductor located on a piezoelectric substrate. 
Two interdigital transducers (IDT)  generate SAWs creating effective dragging potential for the excitations in the superconductor.}
\label{Fig1}
\end{figure}

Here, we consider a 2D superconductor placed on a piezoelectric substrate, thus a Bleustein-Gulyaev SAW~\cite{bleustein1968new, gulyaev1969electroacoustic} creates a longitudinal piezo-potential that drags quasiparticles (Fig.~\ref{Fig1}). 
This acoustic driving induces not only a longitudinal, but also a transverse dc current, emerging without an external magnetic field. 
The microscopic origin of this anomalous response lies in skew scattering -- the asymmetric deflection of quasiparticles by impurities, dictated by the nontrivial topology of the $p$-wave wavefunctions. 

Our results establish the anomalous acoustoelectric effect as a powerful method for diagnosing $p$-wave pairing.
The Hall electric current results in an induced phase difference of the order parameter across the transverse boundaries of the sample. 
This phase shift is not suppressed by electron-hole asymmetry, instead, it is governed entirely by the chiral properties of the condensate, providing a significant advantage for acoustic spectroscopy in probing non-trivial pairing states.

%
%
%
%
%
%


\textcolor{blue}{\textit{Model.---}}
We will focus on the low-frequency regime, where the SAW frequency is much smaller than the SC gap, $\hbar\omega \ll |\Delta|$, suppressing transitions between the condensate and the quasiparticle subsystem. 
If the SAW wave vector satisfies the adiabatic condition $k\xi_0 \ll 1$ with $\xi_0 = \hbar v_F / |\Delta|$ the coherence length, the collective response is governed by the subgap dynamics of the SC and normal components, and the kinetic equation approach provides a natural and powerful framework for analysis~\cite{Aronov01081981}. 
The propagating SAW induces a longitudinal low-frequency electric field ${\mathbf{E}(\mathbf{r},t)}$ with both ${\mathbf{E}(\mathbf{r},t)}$ and ${\mathbf{k}}$ aligned along the surface. 
This field drives the dynamics of both the condensate and quasiparticles. 

The dynamics of the quasiparticle distribution function $f_{\bf p}({\bf r},t)$ and the condensate momentum ${\bf p}_s$ is governed by equations
\begin{gather}\label{AG}\nonumber
\frac{\partial f_{\textbf{p}}}{\partial t}+\frac{\partial \tilde{\varepsilon}_{\textbf{p}}}{\partial\textbf{p}}\frac{\partial f_{\textbf{p}}}{\partial\textbf{r}}-\frac{\partial \tilde{\varepsilon}_{\textbf{p}}}{\partial\textbf{r}}\frac{\partial f_{\textbf{p}}}{\partial\textbf{p}}=-I\{f_{\textbf{p}}\} 
+
\sum_{\bf p'}W^{a}_{\bf p'p}f_{\bf p'}(t), \\
\partial_t {\bf p}_s - \nabla \Phi = e{\bf E},
\end{gather}
%
where the local quasiparticle energy is $\tilde{\epsilon}_{\bf p} = \epsilon_{\bf p} + {\bf v}\cdot{\bf p}_s$, with $\epsilon_{\bf p} = \sqrt{\tilde{\xi}_{\bf p}^2 + \Delta^2}$, $v=p/m$, $\tilde{\xi}_{\bf p} = \xi_{\bf p} + \Phi + p_s^2/2m$, and $\xi_{\bf p} = p^2/2m - E_F$. 
The gauge-invariant potentials are defined as ${\bf p}_s = \frac{1}{2}\nabla\chi - e{\bf A}$ and $\Phi = e\phi + \frac{1}{2}\partial_t\chi$, where $\phi$ and ${\bf A}$ are the scalar and vector potentials, respectively, and $\chi$ is the phase of the order parameter. 
Since the EM field (generated by the SAW) is longitudinal, we choose the ${\bf A}=0$ gauge and treat $\phi$ as the quasistatic potential induced by the SAW on the piezoelectric substrate. 
Furthermore, the collision integral splits into the elastic and inelastic terms, $I\{f\} = I^{\text{el}}\{f\} + I^{\text{in}}\{f\}$, accounting for the impurity scattering and quasiparticle interaction with phonons and other particles. 
The term $W^a_{\mathbf{p}'\mathbf{p}}$ introduces a correction to the elastic component $I^{\text{el}}\{f\}$ to account for asymmetric (skew) scattering processes~\cite{Belinicher1980}. 

We expand the non-equilibrium distribution up to the second order in the acoustoelectric field $\textbf{E}$: $f_{\textbf{p}}\approx n_0(\tilde\varepsilon_{\bp})+f_1(\mathbf{p},\textbf{r},t)+f_2(\textbf{p})$, where $n_0(\tilde\varepsilon_{\bp})$ is equilibrium Fermi-Dirac distribution, $f_1(\mathbf{p},\textbf{r},t)$ is the first-order response, which depends on time and coordinates via the factor $\exp({i\mathbf{k}\cdot\mathbf{r}-i\omega t)}$; $f_2(\bp)$ is stationary part of the second-order response.
In the weakly nonequilibrium regime, the first order correction to quasiparticle distribution function can be decomposed into zero-order and first-order harmonics~\cite{ApenkoLozovik}, and under the assumption $kl\ll 1$ with $l$ the mean free path, $f_1(\mathbf{p},\textbf{r},t) = \delta f_0 + (\mathbf{p} \cdot \mathbf{f}_1)$. 

Regarding the first-order response,  
the kinetic equations take the form (see Supplemental Material~\cite{[{See Supplemental Materials at [URL], which gives the details of the derivations}]SMBG}):
\begin{gather} \label{KineticEqs}
\left(\frac{1}{\tau_\varepsilon} - i\omega\right) \delta f_0 + i \frac{\xi_{\bp}}{\varepsilon_{\bp}} \frac{p^2}{2m} (\mathbf{k} \cdot \mathbf{f}_1) = i\omega \frac{\xi_{\bp}}{\varepsilon_{\bp}} \Phi n'_0, \\
\left( \frac{1}{\tau_{\bp}} - i\omega \right) \mathbf{f}_1 + i \frac{\xi_{\bp}}{\varepsilon_{\bp}} \frac{\mathbf{k}}{m} \delta f_0 = i\omega \frac{\mathbf{p}_s}{m} n_0', \nonumber \\
\omega \mathbf{p}_s + \mathbf{k} \Phi = i e \mathbf{E}, \nonumber
\end{gather}
with a solution
\begin{gather} \label{KineticEqs2}
\delta f_0=\frac{\xi_{\bp}}{\varepsilon_{\bp}}n_0'
\frac{i\omega\Phi+\omega D_{\bp\omega}({\bf k}\cdot{\bf p}_s)}{(1/\tau_\varepsilon-i\omega)+D_{\bp\omega}k^2\xi_{\bp}^2/\varepsilon_{\bp}^2},\\\nonumber
{\bf f}_1=\frac{\omega\tau_{\bp\omega}}{m}n_0'
\frac{\omega{\bf p}_s+{\bf k}\Phi\xi_{\bp}^2/\varepsilon_{\bp}^2}
{(1/\tau_\varepsilon-i\omega)+D_{\bp\omega}k^2\xi_{\bp}^2/\varepsilon_{\bp}^2}.
\end{gather}
Here, $D_{\bp\omega}=v_F^2\tau_{\bp\omega}/2$ is a diffusive coefficient, and the relaxation rates read as 
\begin{subequations}
\begin{equation}
    \frac{1}{\tau_{\bp}} = \frac{1}{\tau}\frac{2\varepsilon_{\bp}^2 + |\Delta|^2}{2|\xi_{\bp}|\varepsilon_{\bp}},
    \label{tau_p}
\end{equation}
\begin{equation}
    \frac{1}{\tau_\varepsilon} = \frac{1}{\tau_{in}} +\frac{1}{\tau}\frac{|\Delta|^2}{|\xi_{\bp}|\varepsilon_{\bp}},
    \label{tau_eps}
\end{equation}
\begin{equation}
    \tau_{\bp\omega} = \frac{\tau_{\bp}}{1-i\omega\tau_{\bp}},
    \label{tau_pw}
\end{equation}
\end{subequations}
%
%
with $\tau$ the energy-independent momentum relaxation time for 2D normal electrons (above $T_c$). 
For energy-dependent scattering (e.g., for the Coulomb centers), the relaxation time can be evaluated at the Fermi level, $\tau \equiv \tau(E_F)$. 
However, the momentum relaxation time $\tau_{\bp}$ in~\eqref{tau_p} has a drastically different energy dependence as compared with the conventional superconductors~\cite{Aronov01081981} (see~\cite{SMBG}). 
Furthermore, the energy relaxation rate~\eqref{tau_eps} consist of two components. 
The inelastic scattering $1/\tau_{in}\ll 1/\tau$, and thus, we can use $1/\tau_{\varepsilon}\sim\Delta^2/\tau$, reflecting the relaxation of particle-hole imbalance.
This impurity-mediated relaxation of particle-hole imbalance is an important feature of unconventional superconductors. 

Next, we can express the current density and the total density fluctuations through the calculated distribution functions: $\mathbf{j} = e N_s \mathbf{p}_s / m + \sum_{\mathbf{p}} \mathbf{v} (\mathbf{p} \cdot \mathbf{f}_1)$ and $\delta N = -\partial_\mu N\Phi + \sum_{\mathbf{p}} \delta f_0 (\xi_{\bp}/\varepsilon_{\bp})$.
Combining these expressions with the continuity equation and the condensate equation of motion allows us to find expressions for the supercurrent and the potential induced by the acoustoelectric field:
%
%
\begin{gather}\label{KineticEqs3}
\Phi_{k\omega}=\frac{-ieV^2_{k\omega}({\bf kE})}{\omega^2-k^2V^2_{k\omega}},\,\,{\bf p}^s_{k\omega}=\frac{ie\omega{\bf E}}{\omega^2-k^2V^2_{k\omega}},
\end{gather}
where
\begin{gather}\label{KineticEqs4}
V^2_{k\omega}=\frac{N_s/m-\sum_{\bf p}n_0'\omega D_{\bp\omega}
\frac{\omega\xi_{\bp}^2/\varepsilon_{\bp}^2-i(1/\tau_\epsilon-i\omega)}
{(1/\tau_\epsilon-i\omega)+ D_{\bp\omega}k^2\xi_{\bp}^2/\varepsilon_{\bp}^2}}
{\partial N/\partial\mu-\sum_{\bf p}\frac{\xi_{\bp}^2}{\varepsilon_{\bp}^2}n_0'\frac{i\omega-D_{\bp\omega}k^2}{(1/\tau_\epsilon-i\omega)+ D_{\bp\omega}k^2\xi_{\bp}^2/\varepsilon_{\bp}^2}}.
\end{gather}

Substituting~\eqref{KineticEqs3} in~\eqref{KineticEqs2} and defining the conductivity via ${\bf j}=\sigma({\bf k},\omega){\bf E}$, we find
\begin{eqnarray}
\label{KineticEqs4}
&&\sigma({\bf k},\omega)=\frac{ie^2\omega}{\omega^2-k^2V^2_{k\omega}}
\Biggl[\frac{N_s}{m}+\\
\nonumber
&&~~~~~+\sum_{\bf p}(n_0')D_{\bp\omega}\frac{\omega^2-k^2V^2_{k\omega}\xi_{\bp}^2/\varepsilon_{\bp}^2}{(1/\tau_\varepsilon-i\omega)+D_{\bp\omega}k^2\xi_{\bp}^2/\varepsilon_{\bp}^2}\Biggr].
\end{eqnarray}
This expression captures the linear electromagnetic response of a 2D superconductor in the diffusive limit.
Importantly, this expression reduces to the conventional expression for the conductivity of a normal-state system above $T_c$~\cite{kittel1963quantum}.

Although we found general expressions describing the first-order response, we can simplify them further using the properties of the SAW, namely, the condition $v_s\ll v_F$ with $v_s=\omega/k$.
The term $V^2_{k\omega}$, entering all the main expressions, in the leading order $V^2_{k\omega}\sim v_F^2\gg v_s^2$, which allows us to neglect the terms proportional to $\textbf{p}_s\sim{\bf E}(v_s/v_F^2)$ and keep only the acoustoelectric potential in the form $\Phi = ie(\textbf{k}\cdot\textbf{E})/k^2$~\cite{Aronov01081981}. 
Furthermore, we consider the scattering rates and the SAW frequency satisfying the usual hierarchy $\tau_{\text{in}}^{-1} \ll \omega \ll \tau_i^{-1}$ providing $\tau_{\bp\omega} \approx \tau_{\bp}$ and allowing us to neglect $1/\tau_{in}$.

The stationary part of the second-order correction to the particle density consists of multiple contributions. 
However, only one contribution appears to provide a non-trivial acoustoelectric current~\cite{SMBG} (which is also the case in conventional superconductors ~\cite{gal1973nonlinear, Aronov1976}):
\begin{subequations}
\label{Eqf2Main01}
\begin{eqnarray}
f_2^{(s)}&=&
-
ie\tau_{\bp}\textbf{k}\cdot\left(\frac{\partial \delta  f_0^{(0)}}{\partial \textbf{p}}\Phi_0^{\ast}-\frac{\partial \delta  f_0^{(0)\ast}}{\partial \textbf{p}}\Phi_0\right)\frac{\xi_{\bp}}{\varepsilon_{\bp}}, \\
f_2^{(a)} &=& e\tau_{\bp}\sum_{\bp'}W^a_{\bp'\bp}f_2^{(s)},
\end{eqnarray}
\end{subequations}
where we introduced complex amplitudes $\Phi_0$ and $ \delta  f_0^{(0)}$ providing $\Phi = \Phi_0e^{i\textbf{k}\textbf{r} - i\omega t} + \Phi_0^\ast e^{-i\textbf{k}\textbf{r} + i\omega t}$, $\delta f = \delta f_0^{(0)}e^{i\textbf{k}\textbf{r} - i\omega t} + \delta f_0^{(0)\ast}e^{-i\textbf{k}\textbf{r} + i\omega t}$, and where $f_2^{(s)}$ and $f_2^{(a)}$ are symmetric and asymmetric second-order corrections to distribution function. 

%
%
%
\begin{figure*}[t!]
    \centering
\includegraphics[width=0.93\columnwidth]{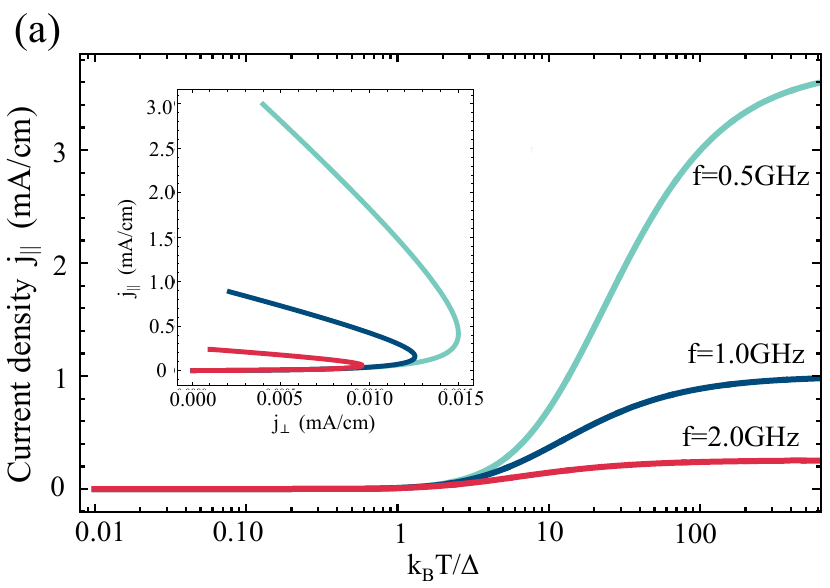} 
\includegraphics[width=0.99\columnwidth]{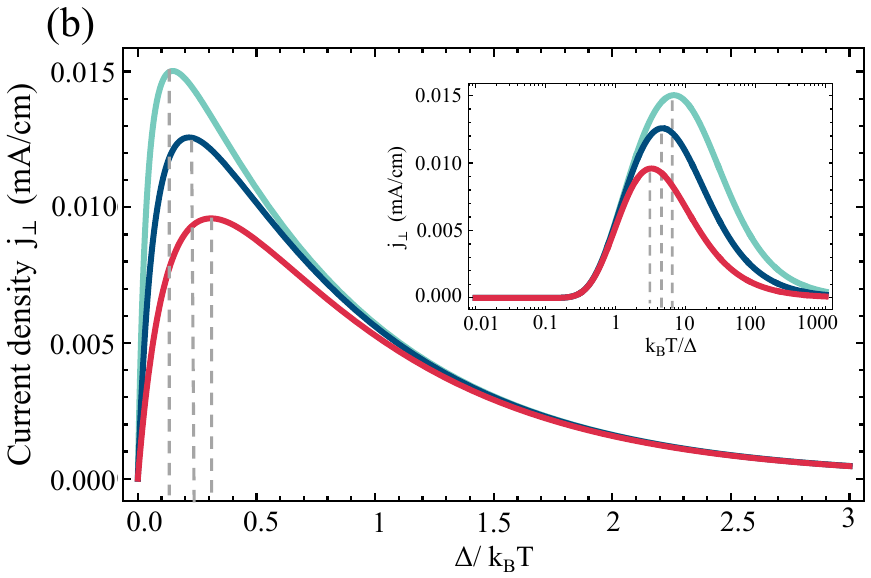} 
\caption
{
(a) Longitudinal component of electric current density $j_{\parallel}$ as a function of dimensionless parameter $k_B T / \Delta$. 
Inset: parametric plot showing the ratio between $j_{\parallel}$ and $j_{\perp}$ for the frequencies $f_{\textrm{SAW}} =$ 0.5, 1, and 2~GHz.
(b) Hall component $j_{\perp}$ as a function of $\Delta / (k_B T)$. 
Dashed vertical lines show the shift of the peak positions. 
Inset: semi-logarithmic scale. 
}
\label{fig1}
\end{figure*}

Given the corrections, the DC electric current reads as
\begin{eqnarray}
\label{EqCurrentMain}
\textbf{j}=e\int \frac{d\textbf{p}}{(2\pi\hbar)^2}\textbf{v}f_2.
\end{eqnarray}
Substituting Eq.~\eqref{Eqf2Main01} in~\eqref{EqCurrentMain}, we find the longitudinal and Hall components of the acoustoelectric current~\cite{SMBG}:
\begin{subequations}
\label{CurrentMain}
\begin{equation}
    \textbf{j}_\parallel = \frac{\textbf{k}}{k^2} \frac{2e^3E^2 \tau}{2\pi\hbar^2}\mathcal{F}(\Delta/k_BT),
    \label{j_par}
\end{equation}
\begin{equation}
    \textbf{j}_\perp = \hat{e}_\perp \frac{2\pi N_iV_0^3 m^2 e^3}{\pi^2 \hbar^7 k}\tau^2 E^2\mathcal{G}(\Delta/k_B T),
    \label{j_perp}
\end{equation}
\end{subequations}
where the functions 
\begin{subequations}
\begin{eqnarray}
    \mathcal{F}(y) &=& 4\int\limits_{y}^\infty dx\left(\frac{x^2 - y^2}{2x^2+y^2}
        \right) \frac{e^{x}}{\left(e^{x}+1\right)^2} \\ \nonumber
        &&\times\frac{\omega(1/\tau_\varepsilon + \mathcal{D}(x,y)k^2)}{\omega^2+(1/\tau_\varepsilon + \mathcal{D}(x,y)k^2)^2},
        \label{F definition}
\end{eqnarray}
\begin{eqnarray}
   \mathcal{G}(y) &=& 4\int\limits_{y}^\infty dx\left(\frac{xy^2\sqrt{x^2 - y^2}}{(2x^2+y^2)^2}
        \right) \frac{e^{x}}{\left(e^{x}+1\right)^2} \\ \nonumber
        &&\times \frac{\omega(1/\tau_\varepsilon + \mathcal{D}(x,y)k^2)}{\omega^2+(1/\tau_\varepsilon + \mathcal{D}(x,y)k^2)^2},
        \label{G definition}
\end{eqnarray}
\end{subequations}
reflect the temperature dependences, and
\begin{equation}
    \mathcal{D}(x,y) = \frac{\tau_{\bp}v_F^2}{2} \frac{\xi^2_{\bp}}{\varepsilon_{\bp}^2} = \tau v_F^2 \frac{(x^2-y^2)^{3/2}}{x(2x^2+y^2)},
\end{equation}
\begin{equation}
    \quad 1/\tau_\epsilon = \frac{1}{\tau}\frac{\Delta^2}{|\xi_{\bp}|\varepsilon_\bp} =\frac{1}{\tau}\frac{y^2}{x\sqrt{x^2-y^2}}.
\end{equation}
Formulas~\eqref{CurrentMain} and below describe the main result of this Letter.

\textit{\textcolor{blue}{Discussion.---}}Figure~\ref{fig1} and its insets show the dependencies of the current densities $\textbf{j}_\parallel$ and $\textbf{j}_\perp$ given by Eq.~\eqref{j_par} and Eq.~\eqref{j_perp} on the parameter $\Delta/(k_B T)$ corresponding to the ratio of the SC gap and temperature (and reversely $k_BT/\Delta$). 
In the limit $T\rightarrow T_c$, the function $\mathcal{F}(\Delta/k_BT)\xrightarrow{T\rightarrow T_c} \omega D_{\bp}k^2/(\omega^2+D_{\bp}^2k^4)$, with $D_{\bp} = \tau_{\bp}v_F^2/2$ being a diffusion coefficient, reflecting the behavior of the longitudinal current component in this limit. 
This yields
\begin{equation}
\textbf{j}_\parallel\xrightarrow{T\rightarrow T_c}\textbf{k} \frac{2e^3E^2 \tau}{2\pi\hbar^2}\frac{\omega D_{\bp}}{\omega^2+D_{\bp}^2k^4},
\end{equation}
which coincides with the longitudinal acoustoelectric current for normal electrons in the limit $kl\ll1$ \cite{PhysRevLett.122.256801}. 
This asymptotics is clearly visible at the right edge of the panel Fig.~\ref{fig1}(a). 
In the opposite limit of zero temperature $k_B T/\Delta \rightarrow 0$ corresponding to the left edge of Fig.~\ref{fig1}(a), the quasiparticles' occupancy above the gap is exponentially suppressed, providing the suppression of the current (Fig.~\ref{fig1}(a)).

The acoustoelectric Hall current vanishes in both the limits $T\rightarrow T_c$ and $T\rightarrow 0$. 
Indeed, at $T\rightarrow T_c$, the material behaves as a normal metal, and the chiral SC order parameter $\Delta$ vanishes, leading to the absence of the skew-scattering terms and the transverse current, which, thus, decays algebraically at $k_BT/\Delta \rightarrow \infty$ (see Fig.~\ref{fig1}(b), inset). 
In the opposite limit $\Delta/k_BT \xrightarrow{T\rightarrow 0} +\infty$, the  current is also exponentially suppressed due to the same reason as for longitudianl current (Fig.~\ref{fig1}(b)). 
Thus, our theory adequately describes the current asymptotics in both the limiting cases $T\rightarrow T_c$ and $T\rightarrow0$. 

Furthermore, the Hall component of the current has maximum at $\Delta \lesssim k_B T$. 
The magnitude and position of the maximum downshifts in terms of $\Delta \lesssim k_B T$ with increasing SAW frequency.
To estimate the current densities, we used parameters typical for MoS$_2$ samples~\cite{parameters}. 
This gives $j^{\textrm{max}}_\perp \approx 0.015$~mA/cm for $f_{\textrm{SAW}} = 0.5$ GHz and $j^{\textrm{max}}_\perp \approx 0.09$~mA/cm for $f_{\textrm{SAW}} = 2$ GHz. 
These magnitudes of the Hall current and the corresponding SC phase differences fall well within an experimentally accessible range.

The longitudinal current has its maxima $j_\parallel \approx 3.3$ mA/cm for $f_{\textrm{SAW}} = 0.5$ GHz and $j_\parallel \approx 0.1$ mA/cm for $f_{\textrm{SAW}} = 2$ GHz at normal phase limit $T = T_c$.
The inset in Fig~\ref{fig1}(b) shows the ratio of the Hall and longitudinal currents for different temperatures and three frequencies of the SAWs. 
For each of the frequencies there is a clear maxima with the highest ratio, which can be important for acquiring larger transverse response of the system.
This very property can be used in detecting the type of superconductivity distinguishing between the $p$-type and other.

\begin{figure}[t!]
    \centering
\includegraphics[width=0.9\columnwidth]{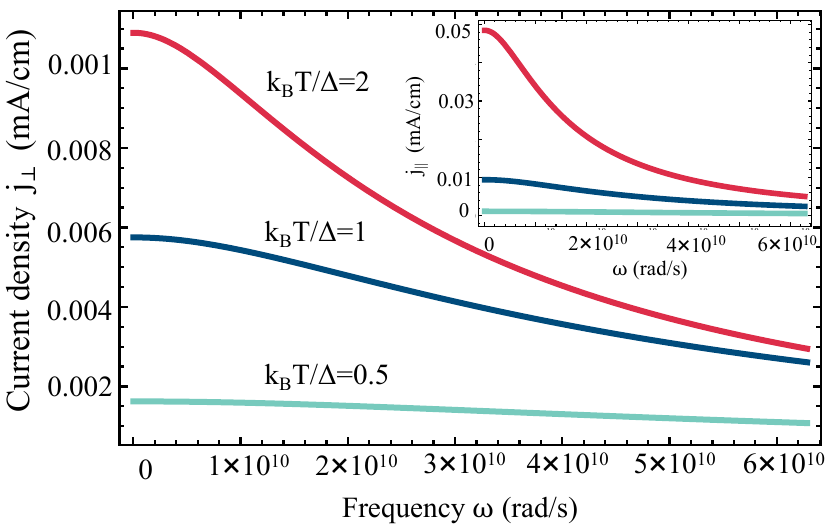}
\caption{\label{Fig3} 
Frequency dependence of the current density components $j_{\perp}$ (main panel) and $j_{\parallel}$ (inset) for several values of the normalized temperature $k_B T / \Delta$. 
The curves correspond to $k_B T / \Delta = 2, 1, \text{ and } 0.5$, respectively.}
    \label{fig:one_column}
\end{figure}

Figure~\ref{Fig3} shows a monotonic decay of both the longitudinal and Hall current densities with increasing SAW frequency. 
To understand this dependence, let us collect the corresponding terms in the normal-state limit, which yields $j\sim[1+ (D_{\bp}k^2/w)]^{-2}$. 
The term $D_{\bp}k^2/w =(\tau_D \omega)^{-1}$ represents a ratio between the acoustic wave period and the characteristic diffusion time $\tau_D = (D_{\bp}k^2)^{-1}$.
Since $k \propto \omega$, the inverse diffusion time $\tau_D^{-1}$ scales quadratically with frequency. 
This result can be interpreted as follows: at higher acoustic frequencies, the diffusion response time becomes shorter. 
Rather than being dragged by the wave, the quasiparticles rapidly diffuse across the shortened wavelength, smearing out the density perturbations and strongly suppressing the macroscopic drag effect.

Furthermore, the most significant attribute of the transverse electric current density is that due to the properties of asymmetric scattering probability~\cite{SMBG}, the nontrivial Hall current appears only for the superconducting order parameter with $p$-wave chiral symmetry, which allows us to detect this type of superconductivity.
Let us note, that in the open-edges geometry (when the sample is free in the transverse direction), the acoustoelectic current of quasiparticles, which we calculated, must be compensated by emerging SC transverse current (the current of Cooper pairs), providing zero total current density:
\begin{equation}
    \textbf{j} +\textbf{j}_s = \textbf{j} + eN_s\textbf{p}_s/m = 0.
\end{equation}
However, in this case, SC condensate acquires a nontrivial transverse phase difference between the sample edges (with the distance $L$ between them), 
\begin{equation}
    |\Delta \phi| = \left|2\int_0^L\frac{p_s}{\hbar}dl\right| = \frac{2L m|j|}{\hbar eN_s},
\end{equation}
which is measurable~\cite{gal1973nonlinear, van1980experimental, parafilo2025proposal, PhysRevLett.39.660, PhysRevLett.122.257001, PhysRevB.25.4515, PhysRevB.108.104506}.

\textcolor{blue}{\textit{Conclusion.---}}
We proposed the anomalous acoustoelectric effect as a robust, non-optical probe of chiral $p$-wave pairing in 2D superconductors.
We demonstrated that a surface acoustic wave induces a transverse dc current in the absence of an external magnetic field, originating from the skew scattering of quasiparticles--a mechanism unique to the chiral pairing symmetry.
Our calculations show that the induced Hall current densities fall well within an experimentally accessible range, on the order of $\mu$A/cm, and predict measurable condensate phase differences across the sample boundaries in open-circuit geometries.
Crucially, near $T_c$, the quasiparticle-mediated response dominates and is not suppressed by the electron-hole asymmetry factor, in stark contrast to the conventional condensate-driven photon drag effect.
These findings establish the anomalous acoustoelectric effect as a powerful electrical method for the experimental identification of unconventional superconductivity, opening new pathways for the discovery of topological phases in layered materials such as transition metal dichalcogenides.


We are grateful to Dr. Anton Parafilo for valuable discussions and Elizaveta Osipova for help with the figures.
We were supported by the Guangdong Basic and Applied Basic Research Foundation under Grant No.~2026A1515012415, National Natural Science Foundation of China (NSFC) under Grant No.~W2532001, 
and the Foundation for the Advancement of Theoretical Physics and Mathematics ``BASIS''. 

\bibliography{biblio}
\bibliographystyle{apsrev4-2}

\end{document}





\title{Supplemental Material for \textit{Anomalous acoustoelectric signatures of chiral superconductivity}}

\author{A.~N.~Osipov}
\affiliation{Technion -- Israel Institute of Technology, Haifa, 3200003, Israel}
\affiliation{Guangdong Technion -- Israel Institute of Technology, 241 Daxue Road, Shantou, Guangdong, China, 515063}

\author{V.~N.~Ivanova}
\affiliation{Guangdong Technion -- Israel Institute of Technology, 241 Daxue Road, Shantou, Guangdong, China, 515063}


\author{V.~M.~Kovalev}
\affiliation{Rzhanov Institute of Semiconductor Physics, Siberian Branch\\ of Russian Academy of Science, Novosibirsk 630090, Russia}
\affiliation{Novosibirsk State Technical University, Novosibirsk 630073, Russia}

\author{I.~G.~Savenko}
\affiliation{Guangdong Technion -- Israel Institute of Technology, 241 Daxue Road, Shantou, Guangdong, China, 515063}
\affiliation{Technion -- Israel Institute of Technology, Haifa, 3200003, Israel}

\date{\today}

\maketitle

\tableofcontents

\section{The kinetic equation}

We start with the Boltzmann kinetic equation:

\begin{gather}
\frac{\partial f_{\textbf{p}}}{\partial t}+\frac{\partial \tilde{\varepsilon}_{\textbf{p}}}{\partial\textbf{p}}\frac{\partial f_{\textbf{p}}}{\partial\textbf{r}}-\frac{\partial \tilde{\varepsilon}_{\textbf{p}}}{\partial\textbf{r}}\frac{\partial f_{\textbf{p}}}{\partial\textbf{p}}=-I\{f_{\textbf{p}}\},
\label{EqBoltzmannSup}
\end{gather}
which is the same as Eq.~(1) from the main text. 
The expansion of the quasiparticles energy reads up to the first order in perturbations:
\begin{equation}
    \tilde \varepsilon_{\bp} \approx \varepsilon_{\bp} + \frac{\xi_{\bp}}{\varepsilon_{\bp}} \Phi + \textbf{v}\textbf{p}_s.
\end{equation}
Then, the terms in the left part of the Boltzmann equation~\eqref{EqBoltzmannSup} transform into:
\begin{subequations}
\begin{eqnarray}
&&\frac{\partial f_{\bf p}}{\partial t}
=
\frac{\partial (n_0+f_1+f_2)}{\partial t}
=
\frac{\partial f_1}{\partial t}
+
n_0'
\frac{\xi_{\bp}}{\varepsilon_{\bp}}e\dot\Phi +n_0' \textbf{v}\cdot\dot{\textbf{p}}_s,  \\
&&\frac{\partial \tilde{\varepsilon}_{\textbf{p}}}{\partial\textbf{p}}\frac{\partial f_{\textbf{p}}}{\partial\textbf{r}}
=
\left[
\frac{\xi_{\bp}}{\varepsilon_{\bp}}\textbf{v}
+\Phi\frac{\Delta^2}{\varepsilon_\mathbf{p}^3}\mathbf{v}
+\textbf{v}_s
\right]
\cdot
\left[
\frac{\partial f_1}{\partial\textbf{r}}
+
n_0'\frac{\xi_{\bp}}{\varepsilon_{\bp}}e
\left(\frac{\partial \Phi}{\partial\textbf{r}}\right) 
+n_0' \textbf{v}\cdot \frac{\partial\textbf{p}_s}{\partial \textbf{r}}
\right],
\\
&&-\frac{\partial {\tilde\varepsilon}_{\textbf{p}}}{\partial\textbf{r}}\frac{\partial f_{\textbf{p}}}{\partial\textbf{p}}
=
\left[-\frac{\xi_{\bp}}{\varepsilon_{\bp}} e\left(\frac{\partial \Phi}{\partial\textbf{r}}\right)
-\frac{\partial(\textbf{v}\cdot \textbf{p}_s)}{\partial \textbf{r}} \right] \left[
\left(\frac{\partial f_1}{\partial\textbf{p}}\right)+
n_0'
\frac{\xi_{\bp}}{\varepsilon_{\bp}}\textbf{v}
+
n_0'
\Phi\frac{\Delta^2}{\varepsilon_{\bp}^3}\mathbf{v}
+ n_0' \textbf{v}_s
\right].
\end{eqnarray}
\end{subequations}


Let us start with the first-order correction, the equation for which has the form:
\begin{eqnarray}
\label{Eqf10}
-i\left(\omega-\textbf{k}\cdot\mathbf{v}\frac{\xi_{\bp}}{\varepsilon_{\bp}}\right)f_1 
-i\omega\frac{\xi_{\bp}}{\varepsilon_{\bp}} \Phi n'_0 -i\omega \textbf{v}\cdot\textbf{p}_sn_0' = I\{f_1\},
\end{eqnarray}
providing the following angular expansion:
\begin{subequations}
\label{angular expansion}
\begin{align}
    &\left(\frac{1}{\tau_\varepsilon} - i\omega\right) \delta f_0 + i \frac{\xi_{\bp}}{\varepsilon_{\bp}} \frac{p^2}{2m} (\mathbf{k} \cdot \mathbf{f}_1) = i\omega \frac{\xi_{\bp}}{\varepsilon_{\bp}} \Phi n'_0, \\
    &\left( \frac{1}{\tau_{\bp}} - i\omega \right) \mathbf{f}_1 + i \frac{\xi_{\bp}}{\varepsilon_{\bp}} \frac{\mathbf{k}}{m} (\delta f_0 +f^{(2)}) = i\omega \frac{\mathbf{p}_s}{m} n_0', \nonumber, \\
    &\left(\frac{1}{\tau_{\bp_2}}-i\omega f_1^{(2)}\right) +ikv \frac{\xi_{\bp}}{\varepsilon_{\bp}}\left(\frac{\bm{k}}{k}\cdot\textbf{f}_1 + f_1^{(3)}\right) = 0,
\end{align}
\end{subequations}
where $f^{(n)} \sim e^{in\phi_{\bp}}$ is the higher-order angular harmonics of the first-order correction. 
Under the assumption of long-wave sound, $kl\ll 1$, $f_1^{(n)} \sim (kl)f_1^{(n-1)}\ll f_1^{(n-1)}$, meaning that the higher-order pump harmonics are responsible for higher-order corrections in terms of small parameter $kl\ll1$, and they can be neglected (leading to Eq.~(2) from the main text).

The equation for the second-order correction in general form reads as:
\begin{eqnarray}
&&i\textbf{k}\cdot\mathbf{v}\frac{\Delta^2}{\varepsilon_{\bp}^3}
\Phi f_1
+ie\textbf{k}\cdot\left(\frac{\partial f_1}{\partial \textbf{p}}\Phi\right)\frac{\xi_{\bp}}{\varepsilon_{\bp}} =
-I\{f_2\}
-
\{I\}_{\Phi}\Phi f_1 - \{I\}_{\textbf{p}_s}\cdot\textbf{p}_s f_1,
\end{eqnarray}
where $\{I\}_{\Phi}$ and $\{I\}_{\textbf{p}_s}$ denote the scattering integrals linearized with respect to the potential and supercurrent, respectively.


\section{Scattering integral for $p$-type superconductor}

Let us consider the collision integral $I\{f_\textbf{p}\}$ for a $p$-type superconductor:
\begin{eqnarray}
\label{EqScatIntGen}
I\{f_\textbf{p}\}=\sum\limits_\mathbf{p'}
\frac{2\pi}{\hbar} N_i|U_{\mathbf{p}\mathbf{p}'}|^2
|  u_\mathbf{p}  u_{\mathbf{p}'}-  v_\mathbf{p}  v_{\mathbf{p}'}e^{i\phi_{\bm{p}}}e^{-i\phi_{\bm{p'}}}|^2
(f_\mathbf{p}-f_{\mathbf{p}'})
\delta( {\varepsilon}_\mathbf{p}- {\varepsilon}_{\mathbf{p}'}),
\end{eqnarray}
where
\begin{eqnarray}
\label{EqUpVptwiddle}
  u^2_\mathbf{p}=\frac{1}{2}\left(1+\frac{ \xi_\mathbf{p}}{\varepsilon_\mathbf{p}}\right),
~~~~
  v^2_\mathbf{p}=\frac{1}{2}\left(1-\frac{ \xi_\mathbf{p}}{\varepsilon_\mathbf{p}}\right)
\end{eqnarray}
are the Bogoliubov coefficients.
We find:
\begin{eqnarray}
\label{EqScatIntGen2}
I\{f_\textbf{p}\} &=&\sum\limits_\mathbf{p'}
\frac{2\pi}{\hbar} N_i|U_{\mathbf{p}\mathbf{p}'}|^2
|u_\mathbf{p} u_{\mathbf{p}'}- v_\mathbf{p} v_{\mathbf{p}'}e^{i\phi_{\bm{p}}}e^{-i\phi_{\bm{p'}}}|^2
(f_\mathbf{p}-f_{\mathbf{p}'})
\delta({\varepsilon}_\mathbf{p}-{\varepsilon}_{\mathbf{p}'}) \\ \nonumber
&=& \sum\limits_\mathbf{p'}
\frac{2\pi}{\hbar} N_i|U_{\mathbf{p}\mathbf{p}'}|^2
|u_\mathbf{p} u_{\mathbf{p}'} - v_\mathbf{p} v_{\mathbf{p}'}|^2
(f_\mathbf{p}-f_{\mathbf{p}'})
\delta({\varepsilon}_\mathbf{p}-{\varepsilon}_{\mathbf{p}'}) 
\\ \nonumber
&&+\frac{2\pi}{\hbar} N_i\sum\limits_\mathbf{p'}|U_{\mathbf{p}\mathbf{p}'}|^2
u_\mathbf{p} u_{\mathbf{p}'} v_\mathbf{p} v_{\mathbf{p}'}(1-\cos(\phi_{\bp,\bp'}))
(f_\mathbf{p}-f_{\mathbf{p}'})
\delta({\varepsilon}_\mathbf{p}-{\varepsilon}_{\mathbf{p}'}) = I_1\{f_\textbf{p}\} + I_2\{f_\textbf{p}\},
\end{eqnarray}
where the first term in the last two lines is the same as for conventional s-type superconductors containing scattering of quasiparticles within one branch~\cite{Aronov01081981}, while the second term depends on angles and contains scattering between quasiparticles from both branches leading to relaxation of particle-hole imbalance~\cite{PhysRevB.92.100506}:
\begin{equation}
    I_1\{\delta f_0\} = 0,\quad I_1\{\textbf{f}_1\} =  \frac{1}{\tau}\frac{|\xi_{\bp}|}{\varepsilon_{\bp}}\textbf{f}_1,
\end{equation}
\begin{eqnarray}
    I_2 &=& \frac{2\pi}{\hbar} N_i\sum\limits_\mathbf{p'}|U_{\mathbf{p}\mathbf{p}'}|^2
2\frac{\Delta^2}{4\varepsilon_{\bp}^2}(1-\cos(\phi_{\bp,\bp'}))
(f_\mathbf{p}-f_{\mathbf{p}'})
\delta({\varepsilon}_\mathbf{p}-{\varepsilon}_{\mathbf{p}'}) \\ \nonumber
&=&  \frac{1}{\tau}\frac{\Delta^2 }{|\xi_{\bp}| \varepsilon_{\bp}}f_{\bp} -\frac{2\pi}{\hbar} (\delta_{n,0}-\frac{1}{2}\delta_{n,1})N_i\left(\int\limits_{0}^{\infty} + \int\limits_{-\infty}^{0}\right)\frac{d\xi_{\bp}}{2\pi\hbar^2}| U_{\mathbf{p}\mathbf{p}'}|^2
2\frac{\Delta^2}{4\varepsilon_{\bp}^2}
f_{\mathbf{p}'}e^{-in\phi_{\bp}}
\delta({\varepsilon}_\mathbf{p}-{\varepsilon}_{\mathbf{p}'})
\\ \nonumber
&=& \frac{1}{\tau}\frac{\Delta^2 }{|\xi_{\bp}| \varepsilon_{\bp}}f_{\bp}
-(\delta_{n,0}-\frac{1}{2}\delta_{n,1})\left(f_{\bp} +\mathcal{P}f_{\bp}\right)\frac{1}{\tau}\frac{\Delta^2}{2|\xi_{\bp}|\varepsilon_{\bp}}.
\end{eqnarray}
Here, $f_\bp\sim e^{in\phi_{\bp}}$ is the $n_{th}$-order angular harmonics, with $\mathcal{P}f_{\bp} = f_{\bp}(\xi_{\bp}\rightarrow -\xi_{\bp})$ being the particle-hole parity transformation and $\tau$ the elastic relaxation time for the normal state. 
Therefore, for a $p$-type superconductor, the relaxation rate depends not only on the angular harmonics of the distribution function, but also on its particle-hole parity. 
Thus, for the odd and even zero and first angular harmonics of the first-order correction (Eq.~(4) in the main text) we have:
\begin{subequations}
    \begin{equation}
        I\{\delta f_0\} +I^{in}\{\delta f_0\} +  = \frac{1}{\tau_{\varepsilon}}\delta f_0 =\left(\frac{1}{\tau_{in}}+ \frac{1}{\tau}\frac{\Delta^2}{|\xi_{\bp}|\varepsilon_{\bp}}\right) \delta f_0,
    \end{equation}
    \begin{equation}
        I\{\textbf{f}_1\} = \frac{1}{\tau_{\bp}}\textbf{f}_1 = \frac{1}{\tau}\frac{2\varepsilon_{\bp}^2 + \Delta^2}{2|\xi_{\bp}|\varepsilon_{\bp}}\textbf{f}_1,
    \end{equation}
\end{subequations}
where for the first-order harmonics we add inelastic relaxations with relaxation time $\tau_{in}\gg\tau$. Let us remark, that only odd functions of energy $f(\xi_{\bp})$ tend to have relaxation on the impurities, meaning that impurities for unconventional superconductors are responsible for the relaxation of particle-hole imbalance.

\section{The asymmetric scattering terms}

The scattering probability reads as
\begin{eqnarray}
\label{ScatteringProb01}
W_{\mathbf{p}'\mathbf{p}}
&=&
\frac{2\pi}{\hbar} N_i
\left|
M_{\mathbf{p}'\mathbf{p}}
+
\sum_\mathbf{l}\frac{M_{\mathbf{p}'\mathbf{l}}M_{\mathbf{l}\mathbf{p}}}{\varepsilon_{\mathbf{p}}-\varepsilon_{\mathbf{l}}+i\delta}
\right|^2\delta( \varepsilon_{\mathbf{p}}- \varepsilon_{\mathbf{p}'})\\ \nonumber
\label{EqMatrix02}
&\approx&
\frac{2\pi}{\hbar} N_i
\left[
|M_{\mathbf{p}'\mathbf{p}}|^2
+
i\pi
\sum_\mathbf{l}
(M_{\mathbf{p}'\mathbf{p}}
M_{\mathbf{p}'\mathbf{l}}^*
M_{\mathbf{l}\mathbf{p}}^*
-
M_{\mathbf{p}'\mathbf{p}}^*
M_{\mathbf{p}'\mathbf{l}}
M_{\mathbf{l}\mathbf{p}})
\delta( \varepsilon_{\mathbf{p}}- \varepsilon_{\mathbf{l}})
\right]
\delta( \varepsilon_{\mathbf{p}}- \varepsilon_{\mathbf{p}'}),
\end{eqnarray}
where the scattering matrix element is
\begin{eqnarray}
M_{\mathbf{p}'\mathbf{p}}&=&
=
V_0
\left(
  u_{\mathbf{p}'}
  u_{\mathbf{p}}
-
  v_{\mathbf{p}'}
  v_{\mathbf{p}}
e^{i\phi_{\mathbf{p}'}}
e^{-i\phi_{\mathbf{p}}}
\right).
\end{eqnarray}
The first term of the scattering probability~\eqref{ScatteringProb01} represents the Born approximation governing elastic scattering on  impurities ($I\{f\}$), which was considered in the previous section. 
The second part governs the skew scattering probability:
\begin{eqnarray}
&&\sum_\mathbf{l}
(M_{\mathbf{p}'\mathbf{p}}
M_{\mathbf{p}'\mathbf{l}}^*
M_{\mathbf{l}\mathbf{p}}^*
-
M_{\mathbf{p}'\mathbf{p}}^*
M_{\mathbf{p}'\mathbf{l}}
M_{\mathbf{l}\mathbf{p}})
\delta(\varepsilon_{\mathbf{p}}-\varepsilon_{\mathbf{l}})\\
\nonumber
&&~~~~~=
V_0^3i\pi
\sum_\mathbf{l}
\left(
 u_{\mathbf{p}'}
 u_{\mathbf{p}}
-
 v_{\mathbf{p}'}
 v_{\mathbf{p}}
e^{i\phi_{\mathbf{p}'}}
e^{-i\phi_{\mathbf{p}}}
\right)
\left(
 u_{\mathbf{p}'}
 u_{\mathbf{l}}
-
 v_{\mathbf{p}'}
 v_{\mathbf{l}}
e^{-i\phi_{\mathbf{p}'}}
e^{i\phi_{\mathbf{l}}}
\right)
\left(
 u_{\mathbf{l}}
 u_{\mathbf{p}}
-
 v_{\mathbf{l}}
 v_{\mathbf{p}}
e^{-i\phi_{\mathbf{l}}}
e^{i\phi_{\mathbf{p}}}
\right)
\delta(\varepsilon_{\mathbf{p}}-\varepsilon_{\mathbf{l}})
\\
\nonumber
&&~~~~~~~
-
V_0^3i\pi
\sum_\mathbf{l}
\left(
 u_{\mathbf{p}'}
 u_{\mathbf{p}}
-
 v_{\mathbf{p}'}
 v_{\mathbf{p}}
e^{-i\phi_{\mathbf{p}'}}
e^{i\phi_{\mathbf{p}}}
\right)
\left(
 u_{\mathbf{p}'}
 u_{\mathbf{l}}
-
 v_{\mathbf{p}'}
 v_{\mathbf{l}}
e^{i\phi_{\mathbf{p}'}}
e^{-i\phi_{\mathbf{l}}}
\right)
\left(
 u_{\mathbf{l}}
 u_{\mathbf{p}}
-
 v_{\mathbf{l}}
 v_{\mathbf{p}}
e^{i\phi_{\mathbf{l}}}
e^{-i\phi_{\mathbf{p}}}
\right)
\delta(\varepsilon_{\mathbf{p}}-\varepsilon_{\mathbf{l}})\\
\nonumber
&&~~~~~=
V_0^3i\pi
\sum_\mathbf{l}
2i u_{\mathbf{p}}^2 v_{\mathbf{p}}^2
\left[
\sin(\phi_{\mathbf{p}}-\phi_{\mathbf{p}'})
+
\sin(\phi_{\mathbf{p}'}-\phi_{\mathbf{l}})
+
\sin(\phi_{\mathbf{l}}-\phi_{\mathbf{p}})
\right]
\delta(\varepsilon_{\mathbf{p}}-\varepsilon_{\mathbf{l}})\\
\nonumber
&&~~~~~=
-V_0^3(2\pi)
\frac{1}{(2\pi)^2}\int\limits_0^\infty ldl\int\limits_0^{2\pi}d\phi_\mathbf{l}
(1-\frac{\xi_{\bp}^2}{\varepsilon_{\bp}^2})
\sin(\phi_{\mathbf{p}}-\phi_{\mathbf{p}'})
\delta(\varepsilon_{\mathbf{p}}-\varepsilon_{\mathbf{l}})\\
\nonumber
&&~~~~~\approx
-V_0^3
\int\limits_{-\infty}^\infty \left(\frac{dl^2}{2m}m-\mu\right)
(1-\frac{\xi_{\bp}^2}{\varepsilon_{\bp}^2})
\sin(\phi_{\mathbf{p}}-\phi_{\mathbf{p}'})
\delta(\varepsilon_{\mathbf{p}}-\varepsilon_{\mathbf{l}}) 
= -2V_0^3 m \int\limits_{0}^\infty d\xi_{l}
(1-\frac{\xi_{\bp}^2}{\varepsilon_{\bp}^2})
\sin(\phi_{\mathbf{p}}-\phi_{\mathbf{p}'})
\delta(\varepsilon_{\mathbf{p}}-\varepsilon_{\mathbf{l}})
\\
\nonumber
&&~~~~~ = -2V_0^3 m \int\limits_{\Delta}^\infty d\varepsilon_{l} \frac{\varepsilon_{\bp}}{|\xi_{\bp}|}
(1-\frac{\xi_{\bp}^2}{\varepsilon_{\bp}^2})
\sin(\phi_{\mathbf{p}}-\phi_{\mathbf{p}'})
\delta(\varepsilon_{\mathbf{p}}-\varepsilon_{\mathbf{l}}) = -2V_0^3 m\frac{\varepsilon_{\bp}}{|\xi_{\bp}|}
(1-\frac{\xi_{\bp}^2}{\varepsilon_{\bp}^2})
\sin(\phi_{\mathbf{p}}-\phi_{\mathbf{p}'}).
\end{eqnarray}
Thus, the scattering probability in the zero order in perturbation potential takes the form:
\begin{equation}
    W_{\bp\bp'}^a\Big|_{\Phi =0} = -\frac{4\pi N_i}{\hbar^3}V_0^3 m{\frac{\varepsilon_{\bp}}{|\xi_{\bp}|}}
(1-\frac{\xi_{\bp}^2}{\varepsilon_{\bp}^2})
\sin(\phi_{\mathbf{p}}-\phi_{\mathbf{p}'}) \delta(\varepsilon_{\bp} - \varepsilon_{\bp'}) = W_0(-2){\frac{\varepsilon_{\bp}}{|\xi_{\bp}|}}
(1-\frac{\xi_{\bp}^2}{\varepsilon_{\bp}^2})
\sin(\phi_{\mathbf{p}}-\phi_{\mathbf{p}'}) \delta(\varepsilon_{\bp} - \varepsilon_{\bp'}),
\end{equation}
with $W_0 = \frac{2\pi N_i}{\hbar^3}V_0^3 m$.


\section{The second-order correction}

The second-order correction has several contributions. 
As long as we are interested in the current density, we can write only the first (transport) harmonic of the second-order distribution function. 
Moreover, we neglect the terms related to $\textbf{p}_s$, due to the reasons discussed in the main text (their smallness). 
Thus, the relevant-to-us part of the second-order correction reads as:
\begin{eqnarray}
\label{Eqf2Main01}
f_2^{(\Phi)}&=&
\left(-\tau_{\bp}\right)
\Bigg\{
i\textbf{k}\cdot\mathbf{v}\frac{\Delta^2}{\varepsilon_{\bp}^3}e
\left(\Phi_0^{\ast}\delta f_0^{(0)}-\Phi_0 \delta f_0^{(0)\ast}\right) +  
ie\textbf{k}\cdot\left(\frac{\partial \delta f_0^{(0)}}{\partial \textbf{p}}\Phi_0^{\ast}-\frac{\partial \delta f_0^{(0)\ast}}{\partial \textbf{p}}\Phi_0\right)\frac{\xi_{\bp}}{\varepsilon_{\bp}}
\\ \nonumber
~~~~~~~~~~~~~~~~~~~
&&+ \{I\}_{\Phi}e \left(\Phi_0^{\ast}(\textbf{k}\cdot\textbf{f}_1^{(0)})+\Phi_0 (\textbf{k}\cdot\textbf{f}_1^{(0)\ast})\right) \Bigg\}.
\end{eqnarray}
These terms provide the following contributions to the electric current density: 
\begin{subequations}
\label{current contributions}
\begin{eqnarray}
\textbf{j}^{(1)}&=&
e\int \frac{d\textbf{p}}{(2\pi\hbar)^2}\frac{\textbf{p}}{m}
\left(-\tau_{\bp}\right)
i\textbf{k}\cdot\mathbf{v}\frac{\Delta^2}{\varepsilon_{\bp}^3}e
\left(\Phi_0^{\ast}\delta f_0^{(0)}-\Phi_0 \delta f_0^{(0)\ast}\right) =0,
\end{eqnarray}
\label{nontriv_current}
\begin{eqnarray}
\textbf{j}^{(2)}&=&
e\int \frac{d\textbf{p}}{(2\pi)^2}\frac{\textbf{p}}{m}
(-\tau_{\bp})\frac{\xi_{\bp}}{\varepsilon_{\bp}}
ie\textbf{k}\cdot\frac{\partial }{\partial \bm{p}}\left( \delta f_0^{(0)}\Phi_0^{\ast}- \delta f_0^{(0)\ast}\Phi_0\right),
\end{eqnarray}
\begin{eqnarray}
\textbf{j}^{(3)}&=& e\int \frac{d\textbf{p}}{(2\pi\hbar)^2}\frac{\textbf{p}}{m}\tau_{\bp}
\{I\}_{\Phi} e\left(\Phi_0^{\ast}(\textbf{k}\cdot\textbf{f}_1^{(0)})+\Phi_0 (\textbf{k}\cdot{\textbf{f}_1^{(0)}}^\ast)\right) =0 .
\label{j3simplified}
\end{eqnarray}
\end{subequations}
%

Here we should give a comment, that, as soon as $\int d\bp = \int d\phi\int\limits_0^{\infty} pdp = m\int d\phi\int\limits_{-\mu}^{\infty} d\xi_{\bp} \approx  m\int d\phi\int\limits_{-\infty}^{\infty} d\xi_{\bp}$, the particle-hole parity (parity in terms of replacement $\xi_{\bp}\rightarrow -\xi_{\bp}$) of the terms is important. 
The term $\delta f_0$ has odd parity. 
Thus, in the first contribution in Eq.~\eqref{current contributions} the expression under the integral is odd, and thus, the integral vanishes.

The third contribution is also zero (for similar reasons): due to the fact that $\textbf{f}_1$ is even, and the lianerization of the scattering integral $\{I^{el}\}_{\Phi}$ is odd. 

These results are consistent with the ones obtained previously for $s$-type superconductors!\cite{Aronov1976, gal1973nonlinear}. 
Thus, only the middle term in Eq.~\eqref{Eqf2Main01} provides a nontrivial contribution to the current. 

Similarly, for the asymmetric part of the distribution function, we only have one  contribution providing non-vanishing current: 
%
\begin{equation}
    -\frac{f_2^{(a1)}}{\tau_{\bp}} = - \sum_{\bp'}W^a_{\bp'\bp}\Big|_{\Phi =0}f_2^{(s)}.
\end{equation}
%

%



\section{The stationary current (DC)}

The stationary part of the current density is given by Eq.~\eqref{nontriv_current}.
That expression can be simplified:
\begin{eqnarray}
\textbf{j}_\parallel&=&
e\int \frac{d\textbf{p}}{(2\pi\hbar)^2}\frac{\textbf{p}}{m}
\left(-\tau_{\bp}\right)\frac{\xi_{\bp}}{\varepsilon_{\bp}}
i\textbf{k}\cdot\frac{\partial }{\partial \bm{p}}\left( \delta f_0^{(0)}\Phi_0^{\ast}- \delta f_0^{(0)\ast}\Phi_0\right) 
=
e\int \frac{d\textbf{p}}{(2\pi\hbar)^2}\frac{\partial }{\partial p_\parallel
}\left(\frac{\textbf{p}}{m}
\left(-\tau_{\bp}\frac{\xi_{\bp}}{\varepsilon_{\bp}}\right)\right)
iek\left(2e\frac{\xi_{\bp}}{\varepsilon_{\bp}}
\frac{E^2}{k^2}
{n_0'} i\mathcal{J}\right) 
\nonumber\\ \nonumber
&=&
e \textbf{k} \int \frac{d\textbf{p}}{(2\pi\hbar)^2}\frac{1}{m}
\left(-\tau_{\bp}\right)\frac{\xi_{\bp}}{\varepsilon_{\bp}}
ie\left(e\frac{\xi_{\bp}}{\varepsilon_{\bp}}
\frac{E^2}{k^2}
2
i\mathcal{J}
{n_0'}\right) + 
{e \textbf{k} \int \frac{d\textbf{p}}{(2\pi\hbar)^2}\frac{\textbf{p}}{m}
\frac{\partial}{\partial\xi_{\bp}}(\tau_{\bp}\frac{\xi_{\bp}}{\varepsilon_{\bp}}) \cos(\phi_{\bp})^2 
iek\left(e\frac{\xi_{\bp}}{\varepsilon_{\bp}}
\frac{E^2}{k^2}
2
i\mathcal{J}
n_0'\right)}
\\ \nonumber
&=&
(i^2)\frac{\textbf{k}}{k^2} \frac{e^3}{2\pi\hbar^2}
\left(2E^2\tau\right)  \int_0^\infty \frac{pd p}{m}
\left(\frac{2|\xi_{\bp}|\varepsilon_{\bp}}{2\varepsilon_{\bp}^2+\Delta^2}\frac{\xi_{\bp}^2}{\varepsilon_{\bp}^2}
\right)\mathcal{J}{n_0'}  
\approx - \frac{\textbf{k}}{k^2} \frac{e^3}{2\pi\hbar^2}
\left(2E^2\tau\right)  2\int_0^\infty d \xi_{\bp}
\left(\frac{2|\xi_{\bp}|\varepsilon_{\bp}}{2\varepsilon_{\bp}^2+\Delta^2}\frac{\xi_{\bp}^2}{\varepsilon_{\bp}^2}
\right) \mathcal{J}(\varepsilon_\bp){n_0'} \\ 
\label{EqWithJ01}
&=&
- \frac{\textbf{k}}{k^2} \frac{2e^3E^2 \tau}{2\pi\hbar^2}
  4\int_\Delta^\infty \left(\frac{\xi_{\bp}^2}{2\varepsilon_{\bp}^2+\Delta^2}
\right) \mathcal{J}(\varepsilon_\bp) d \varepsilon_{\bp}
{n_0'} = -\frac{\textbf{k}}{k^2} \frac{2e^3E^2 \tau}{2\pi\hbar^2}
  4\int_\Delta^\infty \left(\frac{\varepsilon_{\bp}^2 - \Delta^2}{2\varepsilon_{\bp}^2+\Delta^2}
\right)\mathcal{J}(\varepsilon_\bp)  d \varepsilon_{\bp}
{n_0'},
\end{eqnarray}
where
\begin{equation}
    \mathcal{J}(k, \omega, \varepsilon_\bp) = \operatorname{Im}\left[\frac{i\omega}{(1/\tau_\varepsilon - i\omega) +D_pk^2\xi^2_{\bp}/\varepsilon_{\bp}^2}\right] = \frac{\omega(1/\tau_\varepsilon + D_pk^2\xi^2_{\bp}/\varepsilon_{\bp}^2)}{\omega^2+(1/\tau_\varepsilon + D_pk^2\xi^2_{\bp}/\varepsilon_{\bp}^2)^2},
\end{equation}
with $\xi_{\bp}^2 = \Delta^2-\varepsilon_{\bp}^2$.

We can rewrite~\eqref{EqWithJ01} as:
\begin{eqnarray}
\textbf{j}_\parallel&=&
 \frac{\textbf{k}}{k^2} \frac{2e^3E^2 \tau}{2\pi\hbar^2}\mathcal{F}(\Delta/k_BT),
\end{eqnarray}
with
\begin{equation}
    \mathcal{F}(y) = 4\int\limits_{y}^\infty dx\left(\frac{x^2 - y^2}{2x^2+y^2}
        \right) \frac{e^{x}}{\left(e^{x}+1\right)^2} \frac{\omega(1/\tau_\varepsilon + \mathcal{D}(x,y)k^2)}{\omega^2+(1/\tau_\varepsilon + \mathcal{D}(x,y)k^2)^2}.
\end{equation}
Here,
\begin{equation}
    \mathcal{D}(x,y) = \frac{\tau_{\bp}v_F^2}{2} \frac{\xi^2_{\bp}}{\varepsilon_{\bp}^2} = \tau v_F^2 \frac{(x^2-y^2)^{3/2}}{x(2x^2+y^2)}, \quad 1/\tau_\epsilon = 1/\tau_{in} + \frac{1}{\tau}\frac{\Delta^2}{|\xi_{\bp}|\varepsilon_\bp} =1/\tau_{in}+ \frac{1}{\tau}\frac{y^2}{x\sqrt{x^2-y^2}}.
\end{equation}

Next, let us evaluate the transverse component of the current density:
\begin{eqnarray}
    \textbf{j}_\perp &=& e\sum_{\bp}\frac{\textbf{p}_\perp}{m}\tau_{\bp}\sum_{\bp'}W^a_{\bp'\bp}\left(-\tau_{\bp}\right)\frac{\xi_{\bp'}}{\varepsilon_{\bp'}}
ie\textbf{k}\cdot\frac{\partial }{\partial \bm{p}'}\left( \delta f_0^{(0)}\Phi_0^{\ast}- \delta f_0^{(0)\ast}\Phi_0\right) 
\\ \nonumber
&=& 
\hat{e}_\perp eW_0\int \frac{pdp}{2\pi\hbar} \int \frac{d\phi_{\bp}}{2\pi\hbar}\frac{p}{m}\sin(\phi_{\bp})\tau_{\bp}\int \frac{pdp}{2\pi\hbar} \int \frac{d\phi_{\bp}}{2\pi\hbar}\left(-2\frac{\varepsilon_{\bp}}{|\xi_{\bp}|}
(1-\frac{\xi_{\bp}^2}{\varepsilon_{\bp}^2})
\sin(\phi_{\mathbf{p}}-\phi_{\mathbf{p}'})\right)\left(-\tau_{\bp}\right)\frac{\xi_{\bp'}}{\varepsilon_{\bp'}}
iek \\ \nonumber
&&\times \frac{\partial }{\partial p_\parallel'}\left(2e\frac{\xi_{\bp}}{\varepsilon_{\bp}}
\frac{E^2}{k^2}
{n_0'} i\mathcal{J}\right) \delta({\varepsilon_{\mathbf{p}}-\varepsilon_{\mathbf{p}'}})
\\ \nonumber
&=&  \hat{e}_\perp
eW_0\int \frac{pdp}{2\pi\hbar} \int \frac{d\phi_{\bp}}{2\pi\hbar}\frac{p}{m}\sin(\phi_{\bp})\tau_{\bp}\int \frac{pdp}{2\pi\hbar} \int \frac{d\phi_{\bp}}{2\pi\hbar}
\left(2\frac{\varepsilon_{\bp}}{|\xi_{\bp}|}
(1-\frac{\xi_{\bp}^2}{\varepsilon_{\bp}^2})
\frac{\partial }{\partial p_{\parallel}'}\sin(\phi_{\mathbf{p}}-\phi_{\mathbf{p}'})\right)
\left(-\tau_{\bp}\right)\frac{\xi_{\bp'}}{\varepsilon_{\bp'}}
iek\\ \nonumber
&&\times\left(2e\frac{\xi_{\bp}}{\varepsilon_{\bp}}
\frac{E^2}{k^2}
{n_0'} i\mathcal{J}\right) \delta({\varepsilon_{\mathbf{p}}-\varepsilon_{\mathbf{p}'}})
\\ \nonumber
&=& \hat{e}_\perp
\frac{eW_0 m}{2\hbar^2 (2\pi\hbar)^2} \int\limits_{-\infty}^\infty d\xi_{\bp} \tau_{\bp}\int\limits_{-\infty}^\infty d\xi_{\bp'} \left(\frac{\Delta^2}{|\xi_{\bp}|\varepsilon_{\bp}} (-\tau_{\bp}) \frac{\xi_{\bm{p}}}{\varepsilon_{\bp}}iek 
\right)
\left(2e\frac{\xi_{\bp}}{\varepsilon_{\bp}}
\frac{E^2}{k^2}
{n_0'} i\mathcal{J}\right)\delta(\varepsilon_{\mathbf{p}}-\varepsilon_{\mathbf{p}'})
\\ \nonumber
&=& \hat{e}_\perp
\frac{e^3W_0 m}{k(2\pi)^2\hbar^4} \int\limits_{-\infty}^\infty d\xi_{\bp} (\tau_{\bp})^2
2\int\limits_{\Delta}^\infty d\varepsilon_{\bp'}\frac{\varepsilon_{\bp}}{|\xi_{\bp}|} \left(\frac{\Delta^2}{|\xi_{\bp}|\varepsilon_{\bp}} \frac{\xi_{\bm{p}}}{\varepsilon_{\bp}} 
\right)
\left(\frac{\xi_{\bm{p}}}{\varepsilon_{\bp}} E^2
{n_0'}\mathcal{J}\right)\delta(\varepsilon_{\mathbf{p}}-\varepsilon_{\mathbf{p}'})
\\ \nonumber
&=& \hat{e}_\perp
\frac{4e^3W_0 m}{(2\pi)^2 \hbar^4 k} E^2 \int\limits_{\Delta}^\infty d\varepsilon_{\bp} \frac{\varepsilon_{\bp}}{|\xi_{\bp}|}(\tau_{\bp})^2
\frac{\varepsilon_{\bp}}{|\xi_{\bp}|} \left(\frac{\Delta^2}{|\xi_{\bp}|\varepsilon_{\bp}} \frac{\xi_{\bm{p}}}{\varepsilon_{\bp}} 
\right)
\left(\frac{\xi_{\bm{p}}}{\varepsilon_{\bp}}\mathcal{J}
{n_0'} \right) \\
\nonumber
&=& \hat{e}_\perp
\frac{e^3W_0 m}{\pi^2\hbar^4k}\tau^2 E^2 \int\limits_{\Delta}^\infty d\varepsilon_{\bp} \left(\frac{2|\xi_{\bp}|\varepsilon_{\bp}}{2\varepsilon_{\bp}^2 + \Delta^2}\right)^2
\left(\frac{\Delta^2}{|\xi_{\bp}|\varepsilon_{\bp}}
\right)\mathcal{J}
{n_0'}
\\ \nonumber
&=& \hat{e}_\perp
\frac{2\pi N_i m^2 V_0^3 e^3}{\pi^2 \hbar^7 k}\tau^2 E^2 \int\limits_{\Delta}^\infty d\varepsilon_{\bp} \frac{4|\xi_{\bp}|\varepsilon_{\bp}\Delta^2}{(2\varepsilon_{\bp}^2 + \Delta^2)^2} \mathcal{J}
{n_0'} = \hat{e}_\perp
\frac{2\pi N_iV_0^3 m^2 e^3}{\pi^2 \hbar^7 k}\tau^2 E^2 \int\limits_{\Delta}^\infty d\varepsilon_{\bp} \frac{4\Delta^2\varepsilon_{\bp}\sqrt{\varepsilon_{\bp}^2-\Delta^2}}{(2\varepsilon_{\bp}^2 + \Delta^2)^2} \mathcal{J}
{n_0'}
\\ \nonumber
&=& \hat{e}_\perp \frac{2\pi N_iV_0^3 m^2 e^3}{\pi^2 \hbar^7 k}\tau^2 E^2\mathcal{G}(\Delta/k_B T),
\end{eqnarray}
where 
\begin{equation}
    \mathcal{G}(y) = 4\int\limits_{y}^\infty dx\left(\frac{xy^2\sqrt{x^2 - y^2}}{(2x^2+y^2)^2}
        \right) \frac{e^{x}}{\left(e^{x}+1\right)^2} \frac{\omega(1/\tau_\varepsilon + \mathcal{D}(x,y)k^2)}{\omega^2+(1/\tau_\varepsilon + \mathcal{D}(x,y)k^2)^2}.
\end{equation}

To calculate the integrals over the polar angles we use
\begin{equation}
\partial_{p_{\parallel}}\sin(\phi_{\bp}-\phi_{\bp'}) = -\cos(\phi_{\bp}-\phi_{\bp'}) \frac{\partial \phi_{\bp'}}{\partial p_{\parallel}} = -\cos(\phi_{\bp}-\phi_{\bp'}) \frac{\partial \operatorname{arcctg}(p_{\parallel}/p_{\perp})}{\partial p_{\parallel}} = \cos(\phi_{\bp}-\phi_{\bp'})\frac{\sin(\phi_{\bp'})}{p},
\end{equation}
and
\begin{equation}
    \langle\cos(\phi_{\bp}-\phi_{\bp'})\sin(\phi_{\bp'})\sin(\phi_{\bp})\rangle_{\phi_{\bp}, \phi_{\bp'}} = \int\limits_{-\pi}^\pi \frac{d\phi_{\bp'}}{2\pi}\int\limits_{-\pi}^\pi \frac{d\phi_{\bp}}{2\pi}\cos(\phi_{\bp}-\phi_{\bp'})\sin(\phi_{\bp'})\sin(\phi_{\bp}) = \frac{1}{4}.
\end{equation}



\bibliography{biblio}
\bibliographystyle{apsrev4-2}